\documentclass[twocolumn]{svjour3}          
\smartqed  
\usepackage{graphicx}
\usepackage{url}
\usepackage{stmaryrd}
\usepackage{amssymb}

\usepackage[colorlinks=true,
           urlcolor=blue,
           citecolor=blue,
           linkcolor=blue,
           bookmarks=false,
           bookmarksnumbered,
           backref=page,
           pagebackref=page,
           linktocpage=true
           ]{hyperref}

\begin{document}

\newcommand{\code}[1]{\textit{#1}}
\newcommand{\TODO}[2]{\textbf{\textcolor{red}{TODO #1:}} \textcolor{blue}{ #2}} 
\newcommand{\ie}{{i.e.}} 
\newcommand{\eg}{{e.g.}}
\newcommand{\etc}{etc.}
\newcommand{\etal}{\emph{et al.}}

\newcommand{\Kepler}{\textrm{Kepler}}
\newcommand{\PtII}{\textrm{Ptolemy\,II}}

\newcommand{\mypara}[1]{\vspace{8pt}\noindent\textbf{#1}}
\newcommand{\mynote}[1]{\textcolor{red}{[\emph{#1}]}}
\newcommand{\mydef}[1]{\emph{#1}}
\newcommand{\SecRef}[1]{Section~\ref{#1}}
\newcommand{\FigRef}[1]{Figure~\ref{#1}}
\newcommand{\figref}[1]{Fig.\,\ref{#1}}

\newcommand{\dir}[1]{\texttt{#1}}
\newcommand{\moc}[1]{\ensuremath{\mathsf{#1}}}

\newcommand{\ptolemy}{\textsc{Ptolemy\,II}}

\newcommand{\ev}[1]{\ensuremath{\mathsf{#1}}}
\newcommand{\firea}[3]{\ensuremath{#1\stackrel{#2}{\rightsquigarrow}#3}}
\newcommand{\fireb}[3]{\ensuremath{\mathsf{fired}(#1,#2,#3)}}
\newcommand{\run}[2]{\ensuremath{R_{#1{\rightsquigarrow}#2}}}
\newcommand{\trace}[2]{\ensuremath{T_{#1{\rightsquigarrow}#2}}}
\newcommand{\obs}{\ensuremath{\mathcal{O}}}

\newcommand{\firing}[2]{\ensuremath{\ev{F}_{#1}^{#2}}}

\newcommand{\round}[3]{\ensuremath{\ev{R}_{#1}^{#2..#3}}}

\newcommand{\depends}[3]{\ensuremath{#1{\stackrel{#2}{\dashleftarrow}}#3}}

\newcommand{\br}[1]{\ensuremath{\mathcal{#1}}}

\title{Scientific Workflows and Provenance:\\
Introduction and Research Opportunities}



\author{
V\'ictor Cuevas-Vicentt\'in
\and Saumen Dey 
\and Sven K\"ohler 
\and Sean Riddle 
\and Bertram Lud\"ascher
}


\institute{
Dept.\ of Computer Science, University of California, Davis\\
\email{ludaesch@ucdavis.edu} 
}

\date{16 July 2012 / Accepted: 25 July 2012 / \href{http://rd.springer.com/article/10.1007/s13222-012-0100-z}{Published online (Springer)}: 3 October 2012}

\maketitle

\begin{abstract}
Scientific workflows are becoming increasingly popular for compute-intensive and data-intensive scientific applications. The vision and promise of scientific workflows includes rapid, easy workflow design, reuse, scalable execution, and other advantages, e.g., to facilitate ``reproducible science'' through provenance (e.g., data lineage) support. However, as described in the paper, important research challenges remain.  While the database community has studied (business) workflow technologies extensively in the past, most current work in scientific workflows seems to be done outside of the database community, e.g., by practitioners and researchers in the computational sciences and eScience. 
We provide a brief introduction to scientific workflows and provenance, and identify areas and problems that suggest new opportunities for database research.

\end{abstract}

\section{Introduction}
\label{intro}

A scientific workflow is a description of a process for accomplishing a scientific objective, usually expressed in terms of tasks and their dependencies \cite{taylor07:_workf_for_e_scien,ludaescher09:_scien_workf,deelman09:_workf_and_e_scien}.  Scientific workflows aim to accelerate scientific discovery in various ways, e.g., by providing workflow 
\emph{\textbf{A}utomation}, \emph{\textbf{S}caling}, \emph{\textbf{A}daptation} (for reuse), and \emph{\textbf{P}rovenance} support; or ASAP for short.
For the automated execution of repetitive tasks, e.g., \emph{batch processing} a set of files in a source directory to produce a set of output files in a target directory, shell scripts are traditionally used.  Common processing examples include data (re-)formatting, subsetting, cleaning, analysis, etc. Compute-intensive workflows often result from computational science \emph{simulations}, e.g., running climate and ocean models, or other simulations ranging from particle-physics, chemistry, biology, ecology, to astronomy, and cosmology \cite{ludaescher09:_scien_proces_autom_and_workf_manag}. Scientific workflows can be simple, linear chains of tasks, but more complex graph-structured dependencies are also common; e.g., one can think of tools like \texttt{Make} and \texttt{Ant} as special cases of workflow automation \cite{amin2004gridant}.

Workflows need to be \emph{scalable} and \emph{fault-tolerant} to execute reasonably fast in the presence of compute-intensive tasks and ``Big Data''. For example, a \emph{parameter sweep} experiment may require running a program thousands of times with slightly altered input parameters, thus consuming many compute cycles and producing so much data that manually managing it quickly becomes impossible.  Distributed Grid or cloud computing technologies \cite{deelman2004pegasus,wieczorek2005scheduling,CloudStream} and other parallel frameworks such as MapReduce \cite{dean2008mapreduce} and dataflow process networks \cite{lee95dataflow,abramson2008nimrod,vrba2010nornir} can be used to scale workflow execution by exploiting parallel resources.

In addition to optimizing compute cycles, workflows should also save costly ``human cycles'', e.g., by supporting user-friendly workflow designs that are easy to \emph{adapt and reuse}.  Similarly, scientific workflows should encourage modeling at an appropriate level of granularity, so that the underlying experimental process is effectively documented, thus improving communication and collaboration between scientists, and experimental reproducibility.
Many current systems allow to record the \emph{provenance} (i.e., processing history and lineage) of results, and to monitor runtime execution.  Provenance data can then be queried, analyzed, and visualized to gain a deeper understanding of the results, or simply to ``debug'' a workflow or dataset by tracing its lineage back in time through the workflow execution.

In the following, we introduce and describe different aspects and research  challenges of scientific workflows (Section \ref{S:WF}) and provenance (Section~\ref{S:Prov}).  We provide a short summary and conclusions in Section~\ref{S:Conc}.

\section{Scientific Workflows}\label{S:WF}

We give a brief overview of various aspects of scientific workflows, and then summarize related research challenges.  Our terminology and perspective are influenced by our work on \Kepler\ \cite{Kepler}, a scientific workflow system built on top of \PtII\ \cite{TamingHeterogeneityPtolemyApproach}. 

\subsection{Workflow Models of Computation (MoCs)}\label{SS:WF:Mo}

\begin{figure}[t]
\centerline{\includegraphics[width=0.9\columnwidth]{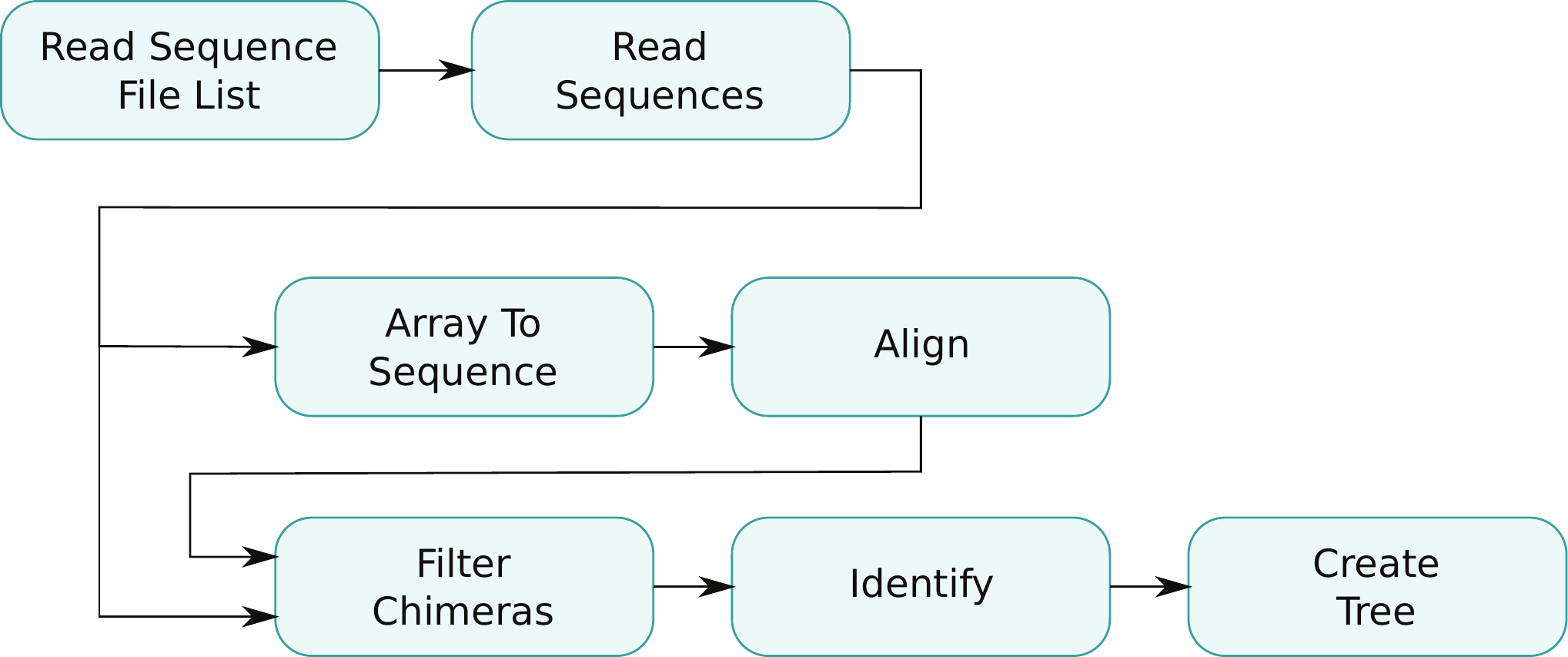}}
\caption{A simplified bioinformatics workflow that reads a list of
lists of genetic sequences:  each sequence is aligned; each
sequence group is then checked for chimeras, which are filtered out; the rest of the group is passed
through. The remaining aligned sequences are identified via a reference
database, and all identified sequences are arranged into a phylogenetic
tree.}
\label{f1:orig_wf}
\end{figure}

\FigRef{f1:orig_wf} depicts a simple example workflow in the form of a dataflow process network \cite{DBLP:conf/ifip/Kahn74,lee08:_seman_of_dataf_with_firin}. Boxes represent \emph{actors} (computational steps) which consume and/or produce \emph{tokens} (data) that are sent over uni-directional \emph{channels} (FIFO queues).  In a process network, actors usually execute as independent, continuous processes, driven by the availability of input tokens.  Alternatively, in some variant models, actors may be \emph{invoked} by a \emph{director} (a kind of scheduler) which coordinates overall execution. Thus, different models of computation (MoCs) can be implemented via different directors \cite{TamingHeterogeneityPtolemyApproach}. 

Let $W$ be a workflow consisting of actors connected through dataflow channels.\footnote{Here we ignore a number of details, \eg, actor
  \emph{ports}, subworkflows ``hidden'' within so-called \emph{composite
    actors}, \etc} With $W$ we can associate a set of \emph{parameters} $\bar p$, input datasets $\bar x$, and output datasets $\bar y$ (not shown in \figref{f1:orig_wf}).  A \emph{model of computation} (MoC) \moc M prescribes how to execute the parameterized workflow $W_{\bar p}$ on $\bar x$ to obtain $\bar y$.  Therefore, we can view a MoC as a mapping $\moc M\,{:}~\mathcal{W} \times \bar P \times \bar X \to \bar Y$ which for any workflow $W\in\mathcal{W}$, parameter settings $\bar p\in\bar P$, and inputs $\bar x\in\bar X$, determines a workflow output $\bar y\in\bar Y$, i.e., $\bar y = \moc{M}(W_{\bar p}(\bar x))$.
Most workflow systems employ a single MoC, while \Kepler\ inherits from \PtII\ several of them (and also adds new ones).

The \dir{PN} (process network) MoC, e.g., is modeled after Kahn process networks \cite{DBLP:conf/ifip/Kahn74}, where each actor executes as an independent, data-driven process.  Thus, channels in \dir{PN} correspond to \emph{unbounded} queues over which ordered token streams are sent, and actors in \dir{PN} \emph{block} (wait) only on read operations, i.e., when not enough tokens are available on input ports.  Process networks naturally support \emph{pipeline-parallelism}
 as well as \emph{task-} and \emph{data-parallelism}.
In the \dir{SDF} (synchronous dataflow) model, each actor has fixed token
consumption and production rates.
This allows the director to  construct a \emph{firing schedule} prior to executing the workflow,  to replace unbounded queues by buffers of fixed size, and  to execute workflows in a single thread, invoking actors according to the static schedule \cite{lee95dataflow}.

Finally, let \dir{DAG} be a MoC that restricts the workflow graph $W$ to a directed, \emph{acyclic} graph of task dependencies, e.g., as in Condor's DAGMan \cite{thain2005distributed}.  In the \dir{DAG} MoC, each actor node $A\in W$ is executed only once,
 and $A$ is executed only after all $A'\in W$
 preceding $A$ (denoted $A'\prec_W A$)  have \emph{finished}
 their execution.
We make no assumption whether $W$ is executed
  sequentially or task-parallel, but only require that any
  \dir{DAG}-compatible schedule for $W$ satisfy the partial order
  $\prec_W$ induced by $W$. A \dir{DAG} director can obtain the legal
  schedules of $\prec_W$ via a topological
  sort of $W$. Finally, note that the \dir{DAG} model can easily
  support task- and data-parallelism, but not pipeline-parallelism.

\paragraph{Research Issues.}

What are suitable MoCs that best satisfy the different requirements of scientific workflows?  Apart from the basic dataflow MoCs above, there are many other formalisms that could be used as foundational models for scientific workflows.  For example, for the Taverna system \cite{Missier:2010:TR:1876037.1876076} a workflow model equivalent to the $\lambda$-calculus is described in \cite{Turi:2007:TWS:1332478.1333522}. For business workflows, Petri nets have been used as a rich and solid foundation, with many theoretical results and practical analysis tools readily available. It is interesting to study to what extent scientific workflows could employ the models and analysis tools developed for business workflows.  Curcin \& Ghanem \cite{curcin2008scientific} ask whether a single system (or a single MoC in our terminology) can cover the requirements of different domains, and conclude 
\begin{quote}
\emph{``... it is highly unlikely that standardization will occur on any one system, as it did with BPEL in the business process domain.''}
\end{quote}
Indeed, the control-oriented modeling common in business process management, and dataflow-oriented modeling in scientific workflows reflect different ways of thinking about workflows \cite{Ludascher:2009:SWB:1617309.1617315,Tan:2009:BSW:1534121.1534133}.

Different formalisms also  imply different modeling and analysis opportunities. In databases, e.g., the relational algebra, relational calculus, and cost-based models yield algebraic, semantic, and cost-based optimization techniques, respectively (pushing selections, query minimization, join-ordering, \etc) Petri net models, on the other hand, allow detailed analysis of concurrent execution behavior, e.g., properties like reachability, liveness, and boundedness. Dataflow networks are amenable to behavioral analysis and verification \cite{lee95dataflow,geilen2003requirements,lee08:_seman_of_dataf_with_firin}. 

 In summary, the quest for suitable models of computation, e.g., to adequately represent computations and to expose and exploit different forms of parallelism, continues. A possible direction are hybrid models \cite{hidders2008dfl,wombacher2010data}, which combine techniques from databases, concurrency models, and stream-processing.

\subsection{Workflow Execution}\label{SS:WF:Exec}

As mentioned in the introduction, workflows need to be scalable to handle compute-intensive and big data loads.  Both implicit and explicit approaches have been used to distribute and parallelize workflow execution.  Consider, e.g., MapReduce \cite{dean2008mapreduce} and its popular open source implementation Hadoop.
 In \cite{wang2009kepler+} an approach is described which allows particular \Kepler\ actors to be distributed onto a Hadoop cluster, i.e., the workflow engine is used for orchestration, but is itself not distributed.  On the other hand, \textsc{Nimrod/k}~\cite{abramson2008nimrod} (also built on top of \Kepler) uses an implicit technique that allows multiple invocations of an actor to execute simultaneously on parallel resources. The approach requires no special configuration to use, but assumes that actors do not maintain state between invocations.
Vrba \etal\ \cite{vrba2009kahn} propose to use Kahn process networks directly to model parallel applications, and argue that this MoC is a flexible alternative to MapReduce. They also report efficiency gains of their framework when comparted to Phoenix, a MapReduce framework specifically optimized for executing on multicore machines.

Recently, the availability of cloud computing has offered new computational resources to many fields in science. Wang \& Altintas \cite{wang12:_early_cloud_exper_with_kepler} report on early experiences with the integration of cloud management and services into a scientific workflow system; Zinn \etal~\cite{CloudStream} propose an approach to support streaming workflows across desktop and cloud platforms.

\paragraph{Research Issues.}

The problem of efficient, scalable workflow execution is intricately linked to the underlying workflow MoC: the more parallelism is exposed by a workflow language and MoC, the more opportunities there are for exploiting it.  An algebraic approach for workflow optimization, well-suited for parameter sweeps, is presented in \cite{ogasawara:hal-00640431}. In essence data are represented by relations, while actors are mapped to operators that either invoke a program or evaluate a relational algebra expression. The semantics of the operators enables workflow optimization by means of rewriting. Analysis and optimization of dataflow process networks \cite{lee95dataflow,lee08:_seman_of_dataf_with_firin} and approaches that combine dataflow, MapReduce, and other parallel techniques with database technologies are also promising \cite{shankar2005integrating,thusoo2009hive,wilde2009parallel,wombacher2010data,borkar2011hyracks}. Last not least, the reemergence of Datalog in real-world, distributed, and workflow applications \cite{hellerstein2010declarative,huang2011datalog,abiteboul2011distributed} presents unique opportunities for database researchers interested in workflows and provenance \cite{dey2012datalog}.

\subsection{Workflow Design}\label{SS:WF:De}

Scientific workflow design shares characteristics with component-based development, serviced-oriented design, and scripting, in that preexisting software components (viewed as \emph{black boxes}) and services are ``glued'', i.e., wired together to form larger applications.  In order to save human cycles, scientific workflow design should be easy and fast, and ideally feel more like storytelling and less like programming. Abstractions like ``boxes-and-arrows'' and flow-charts are often used to develop graphical versions of workflows in a GUI, or to visualize workflow designs, even if they have been specified textually. Depending on the workflow model, different graph formalisms might be used. The simplest designs can be thought of as linear sequences of processing steps, possibly with a stream-processing model like a UNIX pipe. On the other end of the spectrum are complex workflow graphs that can be nested, involve feedback loops, special control-flow elements, etc. For example, Taverna workflows can be nested, have dataflow and control-flow edges, and support a streaming MoC. In addition, Kepler workflows can also have cycles, e.g., to model feedback-loops, or even combine different MoCs when nesting workflows: a top-level \dir{PN} workflow, e.g., may have \dir{SDF} subworkflows \cite{DBLP:conf/iccS/GoderisBALG07}. While there can be practical reasons to employ such sophisticated models \cite{Podhorszki:2007:WAP:1273360.1273368}, complex workflow structures can make it more difficult to adapt and reuse workflows.\footnote{Similarly, in business process modeling, more abstract models, e.g., BPMN, and simple, structured models (e.g., series-parallel graphs) can be easier to understand and  reuse than unstructured or lower-level models, e.g., Petri nets.}

When independently developed, third-party tools and services are combined into workflows and science mash\-ups \cite{howe2009scientific}, the resulting designs can be ``messy'' and may involve complex wiring structures and various forms of software \emph{shims}\footnote{A physical shim is a thin strip of metal for aligning pipes.}, i.e., adapters that transform data so that (part of) the output of one step is made to fit as an input for another step \cite{hull2004treating}.  According to \cite{lin2009task}, a study of 560 scientific workflows from the myExperiment repository \cite{de2009design} showed that over 30\% of workflow tasks are shims.  The proliferation of shims and complex wiring is a limiting factor to wider workflow adoption and reuse.  Such designs are hard to understand (shims distract from the ``real science'') and difficult to maintain: shims and complex wiring make designs ``brittle'', i.e., sensitive to changes in data structures and workflow steps. Complex designs also limit the use of workflows for documenting and communicating the ideas of the experiment, and require expert developers with specialized programming skills, thus increasing the cost for workflow development and maintenance.

BioMoby \cite{BioMoby} aims to improve workflow design by annotating inputs and outputs of actors with semantic type information and enabling the system to provide a wide range of common conversions automatically and transparently. The approach relies on domain experts to create semantic tags and the appropriate conversions.
Wings \cite{gil2007wings} is another system that uses semantic representations to automate various aspects of workflow generation.  The approach described in \cite{bowers2004ontology,bowers2005actor} facilitates service composition and thus scientific workflow design by exploiting structural types, semantic types, and schema constraints between them. Under certain assumptions, schema constraints can be used to automatically generate the desired schema mappings~\cite{fagin2009clio}, allowing scientists to more easily connect workflow components \cite{bowers2004ontology}.  The ``templates and frames'' approach of  \cite{ngu2008flexible} aims at simplifying workflow design via a structured composition of control-flow and dataflow.

\paragraph{Research Issues.}

How can we further simplify workflow design, make workflows less brittle w.r.t.\ change, and thus more easily evolvable and reusable?  How can workflow development become more like storytelling and less like programming?  
The COMAD model (\emph{Collection-Oriented Modeling and Design}), implemented as a  \Kepler\ director \cite{mcphillips06:_collec_orien_scien_workf_integ,mcphillips08:_scien_workf_for_mere_mortal,dou11:_scien_workf_desig}, and the related VDAL model (\emph{Virtual Data Assembly Lines}) \cite{zinn2009scientific,zinn2010xml} use a conveyor-belt assembly-line metaphor to create workflow designs that are mostly linear, and thus easier to understand and modify.
Instead of using shims and complex wiring to ship just the required data fragments to only those actors where they are immediately needed, the conveyor-belt approach instead provides every actor with a subscription mechanism to ``pick-up'' only the relevant parts of the nested (XML-like) data stream for processing. The resulting data is added back to the data stream on the fly.  Every actor thus gets a chance to work on those parts of the data stream that it has subscribed to, leaving other parts untouched. In this way, changes to those other parts, i.e., outside an actor's \emph{read scope}, will not affect that actor's functionality, making the overall approach much more change resilient. As a result, even more so than on a physical assembly line, 
steps can be easily added, swapped, replaced, or removed
in these approaches \cite{mcphillips08:_scien_workf_for_mere_mortal,zinn2010xml,dou11:_scien_workf_desig}:
\begin{figure}[t]
\centerline{\includegraphics[width=0.9\columnwidth]{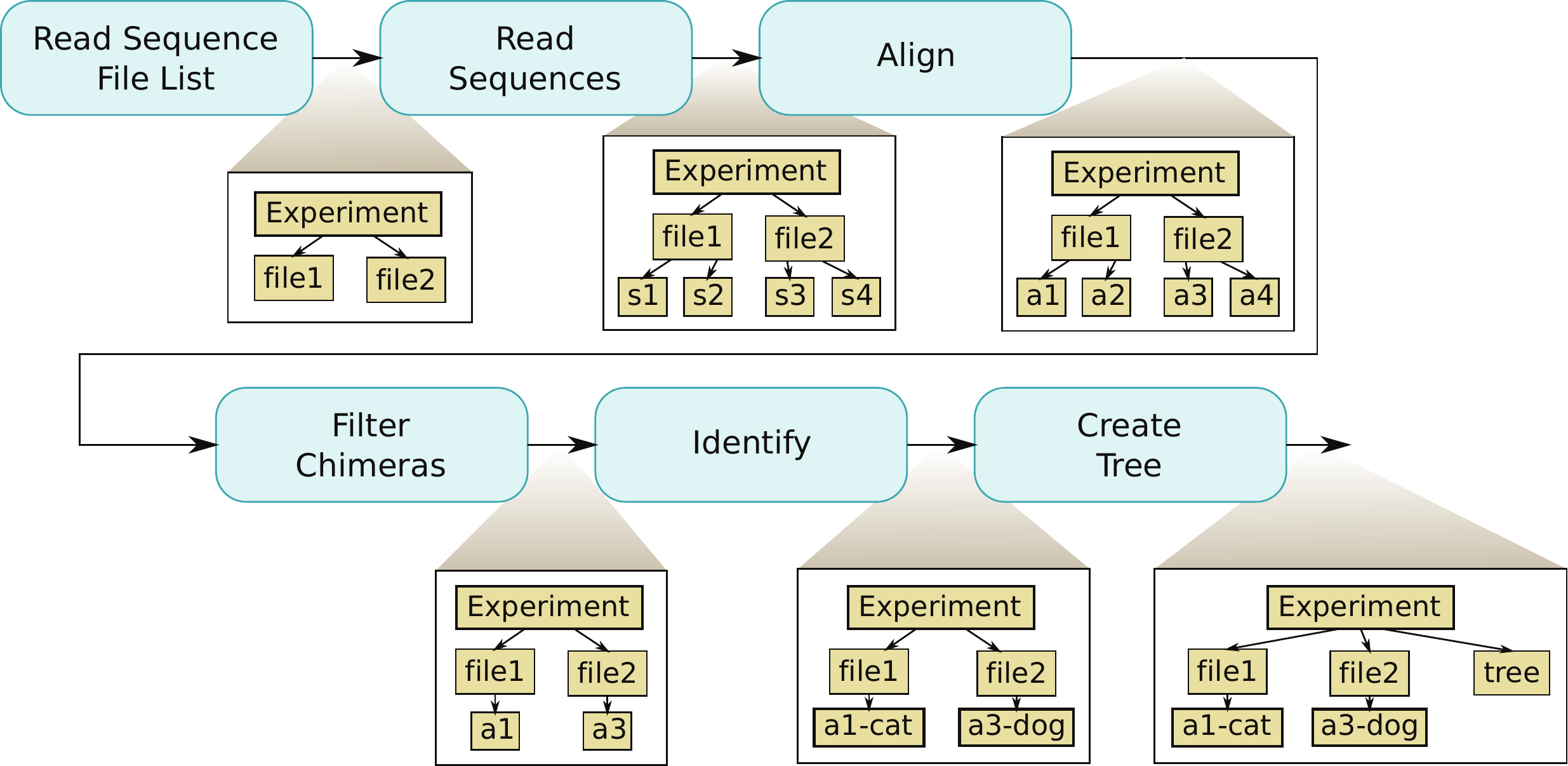}}
\caption{Workflow redesign from \figref{f1:orig_wf} with COMAD/VDAL. Actors can modify parts of the shared data collections that are streamed through the pipeline; \textsf{Read Sequences}, e.g., processes each file entry, replacing it with a collection of the same name
containing all sequences found in the file.
}
\label{f2:comad_wf}
\end{figure}
For example, \FigRef{f2:comad_wf} shows a COMAD redesign of the workflow from \figref{f1:orig_wf}. Note that the shim actor \textsf{ArrayToSequence}\footnote{This shim actor turns a data array token into a sequence of individual data tokens.}
is no longer needed, as the system takes care of the shim problem (a type mismatch, requiring a \texttt{map}-like iteration) using a mechanism based on XPath-like actor configurations (read and write scopes). 
Similarly, the branching has been eliminated, since \textsf{Align} passes data through to \textsf{FilterChimeras}, resulting in a simpler, more user-friendly design.

As another example, in a \emph{curation workflow} \cite{DBLP:journals/procedia/DouCMMLMH12} data and metadata undergo various quality checks (e.g., \emph{Do lat/long-coordinates agree with geo-locations? Is the location spelled right? Are the dates plausible?}), and subsequent clean-up steps: Here, adding, removing, or replacing an actor will not ``break'' the workflow, but only change its curation behavior gracefully. 

In contrast, in  conventional workflows or scripts, complex wiring and control-flow usually prevent simple actor insertions, replacements, or deletions, and workflow changes are much more difficult and error-prone.

These techniques are initial steps towards scientific workflow design for ``mere mortals'' \cite{mcphillips08:_scien_workf_for_mere_mortal}, 
but more work is needed. In COMAD, e.g., users have to configure actors by devising simple queries in the style of XPath or XQuery. Further advances would aim, e.g., at improving data and schema-level support in workflows. Based on structural and semantic information one could develop an ``auto-config'' option for assembly-line workflow designs, where not only shims get absorbed by the system infrastructure, but where actor scope configurations (queries) are automatically inferred. The user-friendly, linear designs currently result in a trade-off between human and compute cycles: all data flows through all ``stations'' (actors) in this workflow model. New analysis techniques could be developed that keep workflow user-views simple, while optimizing executable workflow plans \cite{zinn2009x}. Workflow design technology might also benefit from functional programming ideas, e.g., concatenative, point-free programming and \emph{arrows} \cite{hughes05:_progr_with_arrow}.

\subsection{Workflow Management and Reuse}\label{SS:WF:Man}

Workflow reuse can happen at multiple levels: a scientist may reuse a workflow with different parameters and data, or may modify a workflow to refine the method; workflows can be shared with other scientists conducting similar work, so they provide a means of codifying, sharing, and thus spreading the workflow designer's practice.
A prominent scientific workflow repository is myExperiment \cite{de2009design}, which aims to ``make it easy to find, use and share scientific workflows and other Research Objects, and to build communities''; it currently has more than 5000 members in 250 groups, and a share of over 2000 workflows.\footnote{as of July 2012; see \url{http://www.myexperiment.org}} There is also some work on business process model repositories \cite{Yan:2012:BPM:2109685.2109826}, but the scientific community is more likely to share workflows as a means to accelerate scientific knowledge discovery, whereas the incentive to {publicly} share business processes or ETL workflows is limited by commercial interests \cite{cohen2011search}.

\paragraph{Research Issues.}

Cohen-Boulakia \& Leser argue that \emph{``a wider adoption of SciWFM will only be achieved if the focus of research and development is shifted from methods for developing and running workflows to searching, adapting, and reusing existing workflows''} \cite{cohen2011search}. They propose a number of research directions and problems to better support users of workflows and workflow repositories; e.g., new ways to describe what users search (\emph{workflow sketches}); search for similar (sub-)workflows; and  searching and querying of workflow runs and provenance.
The construction of workflow repositories also entails the development of techniques to deal with the heterogeneity of workflow specifications and metadata, considering aspects such as versioning, view management, configuration, and context management. The development of specific languages for scientific workflows might again borrow ideas from  similar efforts for business processes.  BPQL \cite{Deutch:2012:SQL:2109684.2109809}, e.g.,  enables queries over the structure of a process as well as temporal queries addressing their potential behavior; a graph matching algorithm for workflow similarity search is described in \cite{Dijkman:2009:GMA:1617309.1617317}. Given the different nature and structure of business and scientific workflows \cite{Ludascher:2009:SWB:1617309.1617315}, it is interesting to study how techniques from one area might be adapted or extended for the other.

\section{Provenance and Scientific Workflows}\label{S:Prov}

Provenance is information about the origin, context, derivation, ownership, or history of some artifact \cite{cheney12:_princ_of_proven_dagst_semin}. In the context of scientific workflows a suitable \emph{model of provenance} (MoP) should be based on the underlying model of computation (MoC).  We can derive a MoP from a MoC by taking into account the assumptions that a MoC entails, and by recording the \emph{observables} it affords.  In this way, a MoP captures or at least better approximates real data dependencies for workflows with advanced modeling constructs.  A provenance trace $T$ can be built from a workflow run object $R$, by ignoring irrelevant observables $I$, and adding non-functional observables $M$ that are deemed relevant.  With slight abuse of notation, we can say that $T = R - I + M$, i.e., a trace $T$ (a MoP object), is a ``trimmed'' workflow run $R$ (a MoC object), for which some observables are ignored ($I$) while others (additional metadata) are modeled ($M$). 
More formally, the implementation of a MoC \moc M of a scientific workflow system defines an
\emph{operational semantics}, which in turn can be used to
define a notion of ``natural processing history'' or \emph{run}
\run{\bar x}{\bar y} for $\bar y = \moc M(W_{\bar p}(\bar x))$. 
Given a run \run{\bar x}{\bar y},
the semantics given by \moc M allows us to check
whether \run{\bar x}{\bar y} is a ``faithful'' (or \emph{legal})
execution of $\bar y = \moc M(W_{\bar p}(\bar x))$. For example, the order
of actor firings in a legal run \run{\bar x}{\bar y} must conform to
$\prec_W$; and $\bar p, \bar x$, and $\bar y$ must appear in \run{\bar
  x}{\bar y} as inputs and outputs, respectively.

\paragraph{Observables.} The records of a run \run{\bar x}{\bar y} of a
workflow execution of $\bar y = M(W_{\bar p}(\bar x))$ are built from
basic \emph{observables} associated with \moc M. For $\moc M=\dir{DAG}$, the
observables of a run are the single \emph{firings} of an actor $A$,
together with the inputs $d$ and outputs $d'$ of the firing, recorded
as \fireb d A {d'}.  We may dissect firing events into smaller
observables, \eg, into two records of the form
$\ev{received}(A,m_1(d))$ and $\ev{sent}(A,m_2(d'))$, and a third
record of the form $\ev{caused}(m_1, m_2)$.
This can be useful for MoCs such as \dir{CSP} (\emph{Communicating Sequential Processes}) or \dir{MPI} (\emph{Message Passing Interface}), where actors react to different messages, not just the read (input/token consumption) and write (output/token production) that we used here for \dir{DAG}.  Similarly,  job-based (or \emph{Grid}) workflow systems are usually based on the \dir{DAG} MoC. For these, \fireb d A {d'} may be modeled differently, \eg, based on a job's \ev{start} and \ev{finish} events.

\subsection{Models of Provenance (MoPs)}\label{S:Prov:Mod}

The models for provenance used in scientific workflow systems include custom models such as the RWS (read, write, state-reset) model \cite{Bowers:2006:MUD:2165554.2165573}, and community efforts such as the Open Provenance Model (OPM) \cite{Moreau:2011:OPM:1967762.1967931}. OPM was developed as a minimal, generic standard for provenance, not just for workflows.  In OPM, events are re\-corded when actors consume tokens (\code{used} events) and produce tokens (\code{wasGeneratedBy} events). Thus, storing provenance data can effectively make persistent the data tokens that have been flowing across workflow channels, either through the actual data or by handlers.

Event records typically include an identifier for the actor (e.g., a service or program). The specific port or parameter of the actor associated with the data token can also be recorded, which corresponds to the \code{role} in OPM's \code{usedBy} and \code{generatedBy} events. Each event is generally recorded along with its timestamp. While OPM only considers processes, additional information can be recorded like the notion of state-reset events and rounds in RWS, which define logical units of work. The collection of event records defines a directed acyclic graph (DAG) that represents the execution history.

The OPM also defines \code{wasTriggeredBy} and \code{wasDerivedFrom} relationships. The former denotes that an instance of an actor execution is causally linked to a preceding actor execution, while the later denotes that a data artifact results (at least partially) from processing an earlier artifact. Under an extended interpretation of the OPM \code{used} and \code{wasGeneratedBy} relations, \code{wasTriggeredBy} and \code{wasDerivedFrom} relations can be inferred, which is not always the case in terms of the original standard. The \code{agents} in the OPM, correspond in scientific workflow systems either to software entities that execute workflow components, or to users that initiate and monitor the execution of a workflow. Both cases can be represented in OPM by the \code{wasControlledBy} relation.

\paragraph{Research Issues.}

While OPM and its W3C successor PROV are gaining popularity, by design they leave out specifics of MoCs and custom MoPs. Elements specific to workflows are not present, like firing constraints \cite{dey2012datalog} or the workflow structure.  OPM's temporal semantics is somewhat ambiguous as pointed out in \cite{Moreau:FOPM09}.  An interesting area of research is the development of richer provenance models, corresponding to different workflows systems and MoCs.  For example, the DataONE\footnote{\url{http://www.dataone.org/}} Provenance Working Group develops a unified MoP reflecting scientific workflows provenance from different systems (initially: Kepler, Taverna, VisTrails, and R).

\subsection{Capturing Provenance}\label{S:Prov:Cap}

Provenance in workflows is not limited to the execution of a fixed workflow, but can also include the history of the workflow design, i.e., \emph{workflow evolution provenance}. VisTrails \cite{Freire:2006:MRS:2165554.2165557} keeps track of all the changes that have led to new versions of a given workflow.
Formally, a \code{vistrail} is a tree in which nodes represent workflow versions and edges correspond to operations in a change algebra, such as \code{addConnection} or \code{addModule}.

Difficulties in capturing provenance arise in practice, as scientific workflow systems are built on and interoperate with other systems, e.g., databases, parallel computing platforms, web services, scripting languages, etc. Provenance data originating from lower-level components needs to be made accessible to the workflow system, e.g., resource usage statistics, failures, and repeated execution attempts in parallel programs. Swift \cite{GadelhaJr.:2011:PMS:1967762.1967947} captures such information from high level SwiftScript programs by means of wrapping scripts running in the background.  Each level of abstraction associated with a software layer may include different provenance observables: e.g., the PASS system supports provenance at the file system, workflow engine, script language, and browser levels \cite{Muniswamy-Reddy:2009:LPS:1855807.1855817}.

\paragraph{Research Issues.}

It is desirable to keep the overhead for capturing provenance to a minimum.  An interesting possibility is to study declarative, domain-specific language for provenance in computing systems. This will allow the user to define relevant provenance information at the desired granularity.

With the use of cloud computing and key-value stores for scientific computing, it becomes challenging to capture the necessary provenance data, since these platforms can only be accessed through interfaces.\footnote{See, for example, Amazon's Simple Storage Service (S3) \url{http://aws.amazon.com} and Simple Workflow Service (SWS)} 
With respect to recording overhead, data poses the greatest challenge, since in some cases the size of provenance data can exceed the combined input and output data \cite{Chapman:2008:EPS:1376616.1376715}. Thus, with large-scale data it is essential to identify the information to record and to employ efficient data management techniques. 

\subsection{Storage and Querying}\label{S:Prov:Store}

Scientists want to use provenance data to answer questions such as: \textit{Which data items were involved in the generation of a given partial result?} or \textit{Did this actor employ outputs from one of these two other actors?} Such questions addressed over the provenance data represented by a directed graph translate into two well-known types of directed graph queries: reachability and regular path queries.

A \emph{reachability query} over a directed graph $G$ returns pairs $(x,y)$ of nodes, connected by a path in $G$.
The paths themselves are often not computed.
In addition  to simple graph-traversal and transitive closure algorithms, specialized techniques can be used, e.g., using simpler structures such as chains or trees to compute and compress the transitive closure. In \cite{Jin:2011:PER:1929934.1929941}, a \emph{path-tree} is introduced along with additional techniques to yield a more efficient solution. An approach specifically for provenance graphs is introduced in \cite{Bao:2010:OLS:1807167.1807244}: a labeling scheme captures parallel instantiations (forks) along with loops; labels are of logarithmic length and generated in linear time.

Reachability queries can be computed by RDBMSs with simple extensions.  While \emph{transitive closure} is not a first-order query, it is still a \emph{maintainable} relation through first-order (plus aggregation for some cases) queries \cite{Dong99maintainingtransitive}. This database technique is used, e.g., in the Swift system \cite{GadelhaJr.:2011:PMS:1967762.1967947}. A recent extension of Swift \cite{gadelha2011provenance} now uses a SQL function with a \texttt{RECURSIVE} clause.

An extended form of reachability query is presented in \cite{Missier:2010:FEL:1739041.1739079}, where the authors consider fine-grained provenance data resulting from Taverna workflows that also operate on lists. Their approach is based on using the workflow specification as an index. This is particularly useful to reduce search time for focused queries, where users are interested only in selecting the inputs related to one specific output.

A \emph{regular path query} (RPQ)  over an edge-labeled, directed graph $G$ returns all pairs of nodes $(x,y)$ which are connected in $G$ via a path $\pi$ whose labels spell a word that matches a given regular path expression~$R$.
Regular path expressions are built similar to regular expressions (e.g., concatenation $R_1{\cdot} R_2$, alternation $R_1{\mid} R_2$, Kleene-star $R^*$, etc.) but can also include graph-specific extensions such as $R^{-1}$, denoting edge reversal.

The evaluation of RPQs via simple paths (i.e., paths without repeated nodes) is NP-complete \cite{Mendelzon:1995:FRS:219375.219400} in general, but drops to PTIME when allowing non-simple paths as well. 
The traditional approach to evaluate RPQs is to view the graph as a non-deterministic finite automaton, then construct an equivalent deterministic finite automaton to be used as an (often very large) index.
To achieve a more efficient evaluation, a heuristic approach is presented in \cite{RPQLG}, based on finding rare labels and using them as starting points for bi-directional searches. Each search corresponds to a sub-query, and the combination of results yields the result for the original query. The fact that rare labels are chosen restricts the search significantly and thus the overall computation required.

An alternative way to evaluate RPQs is the translation to Datalog queries. 
A direct translation is given in \cite{Afrati97chainqueries}, whereas \cite{Tekle:2010:GQT:1836089.1836093} provides another, optimized method.

Query languages able to express RPQs include those originally developed for semistructured data integration systems such as STRUDEL \cite{Fernandez:2000:DSW:765218.765222} and proposed extensions on SPARQL. Such languages and other alternatives are discussed in greater depth in \cite{Wood:2012:QLG:2206869.2206879}. However, graph queries issued directly against physical data representations (e.g. XML or RDF) can be difficult to express and expensive to evaluate. In \cite{Anand:2010:TEQ:1739041.1739078} QLP, a high-level query language for provenance graph queries is introduced to address these shortcomings. It provides constructs for querying both structure and lineage information complemented with optimization and lineage-graph reduction techniques.

Visual querying is also well suited for provenance data. In particular, VisTrails \cite{Freire:2006:MRS:2165554.2165557} includes a visual query-by-example interface; additionally, it can perform keyword search on provenance data.

\paragraph{Research Issues.}
Efficient storage and querying of large amounts of provenance data remain important areas of research, in particular, as more provenance data is being produced and shared.  The use of information retrieval techniques in databases has received significant attention (e.g. \cite{Li:2011:PBK:1938123.1938150}), their application specifically to provenance data represents another interesting research direction. Provenance data can also be seen as (detailed) variants of event logs of information systems, thus providing interesting new areas and applications of process mining \cite{vanderAalst:2011:PMD:1983446}. This approach could allow discovering a workflow from provenance data, monitoring its conformance to the expected behavior, or improving the workflow by identifying, e.g., opportunities for optimization and parallelization.

\subsection{Interoperability}\label{S:Prov:Interop}

Today's scientific
experiments are very complex, involving multiple teams working
in collaboration and using various scientific workflow
systems to design and execute respective workflows. In this collaborative
setting, an output data product of a workflow is often used as an input data product of another workflow. Thus, to completely understand a data product we need its provenance information along with provenance information of all the dependent data products. Thus, we need to integrate all the provenance information to effectively answer all provenance queries. 

\paragraph{Heterogeneous MoPs.} Most scientific workflow
systems provide a method for recording provenance. However, these systems use specific provenance
and storage models. For example, Kepler records OPM-based provenance into a
relational database, whereas COMAD employs its own provenance model and stores
provenance information into an XML file. 
In order to answer provenance queries, we need to develop methods to integrate provenance information from various MoPs. A mediation-based approach to solve this problem is presented in \cite{ellqvist2009using}, whereas in \cite{5671861} a framework and common data model for traces is proposed.

\paragraph{Lack of Shared Data Identification.} While working
collaboratively, a scientist may copy a data product into his local system and then perform some formatting tasks before using it as an input data product. The copying and formatting tasks may give
rise to a different identifier for the same data, as the
identification management is being done by individual systems in isolation.
Thus, it is important to identify
these copies so that they are linked appropriately and correct dependencies are established. Missier \etal\ \cite{missier2010linking}
observe this problem and provide a prototypical foundation toward solving it.

\subsection{Provenance Applications}\label{S:Prov:App}

\paragraph{Privacy-aware Provenance.}

While provenance information is very useful, it often carries sensitive information causing privacy concerns, which can be tied to data, processes, and workflow specifications. It is possible to infer the value of a data product, the functionality (being able to guess the output of an actor given a set of inputs) of a process, or the execution flow of the workflow. In \cite{davidson2010privacy} a mathematical foundation to achieve $\varepsilon$-privacy for a process or a workflow is developed. This model is able to compute the input/output combinations for which the required level of $\varepsilon$-privacy is achieved.

SecurityViews \cite{chebotko2008scientific} are developed in order to provide a partial view of the workflow through a role-based access control mechanism, and by defining a set of access permissions on processes, channels, and input/output ports as specified by the workflow owner at design time. 

Two techniques are developed in \cite{dey2011propub} to customize the provenance information based on scientists privacy and data sharing requirements. This approach adheres to a provenance model in which the scientist wants to share the customized provenance data.

While trying to honor the privacy concerns, current techniques remove the private information without providing clear guarantees of 
what queries could be answered using the customized provenance. Also, as the provenance information often is very large, it is important to develop techniques so that privacy and publication requirements can be expressed at a higher level (for e.g. workflow specification, data collection, etc).

\paragraph{Information Overload.}

Provenance data captured by executing a workflow is often larger than the size the actual data \cite{braun2006issues}. ZOOM*UserViews \cite{biton2007zoom} provides a partial, zoomed-out view of a workflow, based on the user-defined distinction between relevant and irrelevant actors. Provenance information is captured based on this view. Another approach in \cite{dey2011propub} does not impose any restriction in capturing provenance information. It allows scientists to specify the customization requests on the provenance information to remove the irrelevant parts, which allows scientists to better understand the relevant parts of the provenance information.

The volume explosion of provenance information gives rise to two very challenging issues: how to effectively and efficiently (i) visualize and browse provenance information under different levels of abstraction, and (ii) specify what are the relevant portions of the provenance information.

\paragraph{Debugging.} 

Provenance recording also enables a natural way of debugging workflows. For instance, data values can easily be inspected and checked for correctness. In addition, by comparing a workflow description with a trace of its run, actors that never fire can easily be detected. 
For this purpose, a static analysis technique is described in \cite{APG} to infer an abstract provenance graph (APG) from a VDAL style workflow description described earlier. The APG allows identifying incorrect configurations and actors that are never fired.

An interesting research issue is how to use provenance data potentially including time-stamps to analyze the efficiency of a workflow execution. 
Independent subworkflow executions would be easily identifiable from data dependencies in a trace graph. 
Those independent subworkflows should optimally be executed in parallel and time-stamps in the trace would allow finding places with suboptimal scheduling.

\paragraph{Fault Tolerance.} 

Similarly to database recovery using log files, provenance can be used to efficiently recover faulty workflow executions as shown in \cite{FT}. 
The trace contains all data items processed and created before the fault as well as the status of invocations at the time of the fault.
Thus, successfully completed invocations can be skipped and data that was in the intra-actor queues at the time of the fault is restored.
The challenges in developing recovery techniques are that
(1) actors can be invoked multiple times and maintain state from one invocation to the next; 
(2) data can be transported between actors outside of the queue-based infrastructure, circumventing provenance recording;
(3) non-trivial scheduling algorithms are used for multiple actor invocations, which are based on data availability.

Some MoCs such as COMAD use complex data structures which are
exchanged between actor invocations in fragments and therefore implicit
dependencies between data artifacts exist that have to be maintained.
This requires special consideration during the recovery process. 
Furthermore, the stateful layer in each actor handling the scope matching
requires new techniques in order to allow an efficient recovery.

\section{Summary and Conclusions}\label{S:Conc}

Scientific workflow systems can help scientists design and execute computational experiments efficiently, but many research issues remain to be solved.  We have given an introduction and overview on scientific workflows and provenance and highlighted a number of research areas and problems.  \emph{Business} workflows (and business process modeling) have been and are being studied extensively by the database community.  With this paper we hope to help trigger or reignite interest in the database community to address some of the challenges in \emph{scientific} workflows. After all, database researchers were among the first to explore the challenges in scientific workflows and to apply database technologies towards them \cite{wainer96:_scien_workf_system,ailamaki1998scientific}. In the era of data-driven scientific discovery and Big Data, there was probably never a better time for researchers to embrace the challenges and opportunities in scientific workflows and to advance the state-of-the-art in data-intensive computing.

\begin{acknowledgements}
  Work supported in part by NSF awards OCI-0830944, OCI-0722079, DGE-0841297, and DBI-0960535.

\end{acknowledgements}

%
%
\bibliographystyle{spmpsci}      
\bibliography{db-sciwf-prov-main}   

\begin{thebibliography}{10}
\providecommand{\url}[1]{{#1}}
\providecommand{\urlprefix}{URL }
\expandafter\ifx\csname urlstyle\endcsname\relax
  \providecommand{\doi}[1]{DOI~\discretionary{}{}{}#1}\else
  \providecommand{\doi}{DOI~\discretionary{}{}{}\begingroup
  \urlstyle{rm}\Url}\fi

\bibitem{vanderAalst:2011:PMD:1983446}
van~der Aalst, W.M.P.: Process Mining: Discovery, Conformance and Enhancement
  of Business Processes, 1st edn.
\newblock Springer (2011)

\bibitem{abiteboul2011distributed}
Abiteboul, S., Bienvenu, M., Galland, A., Rousset, M.: Distributed datalog
  revisited.
\newblock Datalog Reloaded pp. 252--261 (2011)

\bibitem{abramson2008nimrod}
Abramson, D., Enticott, C., Altinas, I.: Nimrod/{K}: Towards Massively Parallel
  Dynamic Grid Workflows.
\newblock In: Supercomputing Conference. IEEE (2008)

\bibitem{Afrati97chainqueries}
Afrati, F., Toni, F.: Chain queries expressible by linear datalog programs.
\newblock In: Deductive Databases and Logic Programming (DDLP), pp. 49--58
  (1997)

\bibitem{ailamaki1998scientific}
Ailamaki, A., Ioannidis, Y., Livny, M.: Scientific workflow management by
  database management.
\newblock In: Intl.\ Conf.\ on Scientific and Statistical Database Management
  (SSDBM), pp. 190--199 (1998)

\bibitem{amin2004gridant}
Amin, K., von Laszewski, G., Hategan, M., Zaluzec, N., Hampton, S., Rossi, A.:
  GridAnt: A Client-Controllable Grid Workflow System.
\newblock In: Hawaii Intl.\ Conf.\ on System Sciences (HICSS). IEEE (2004)

\bibitem{Anand:2010:TEQ:1739041.1739078}
Anand, M.K., Bowers, S., Lud\"{a}scher, B.: Techniques for efficiently querying
  scientific workflow provenance graphs.
\newblock In: Proceedings of the 13th International Conference on Extending
  Database Technology, EDBT '10, pp. 287--298. ACM, New York, NY, USA (2010)

\bibitem{Bao:2010:OLS:1807167.1807244}
Bao, Z., Davidson, S.B., Khanna, S., Roy, S.: An optimal labeling scheme for
  workflow provenance using skeleton labels.
\newblock In: SIGMOD, pp. 711--722 (2010)

\bibitem{biton2007zoom}
Biton, O., Cohen-Boulakia, S., Davidson, S.: {Zoom* userviews: Querying
  relevant provenance in workflow systems}.
\newblock In: VLDB, pp. 1366--1369 (2007)

\bibitem{borkar2011hyracks}
Borkar, V., Carey, M., Grover, R., Onose, N., Vernica, R.: Hyracks: A flexible
  and extensible foundation for data-intensive computing.
\newblock In: Intl.\ Conf.\ on Data Engineering (ICDE) (2011)

\bibitem{bowers2004ontology}
Bowers, S., Lud{\"a}scher, B.: An ontology-driven framework for data
  transformation in scientific workflows.
\newblock In: Data Integration in the Life Sciences (DILS), pp. 1--16 (2004)

\bibitem{bowers2005actor}
Bowers, S., Lud{\"a}scher, B.: Actor-oriented design of scientific workflows.
\newblock Conceptual Modeling (ER) pp. 369--384 (2005)

\bibitem{Bowers:2006:MUD:2165554.2165573}
Bowers, S., Mc{P}hillips, T., Lud\"{a}scher, B., Cohen, S., Davidson, S.B.: A
  model for user-oriented data provenance in pipelined scientific workflows.
\newblock In: Intl.\ Provenance and Annotation Workshop (IPAW) (2006)

\bibitem{braun2006issues}
Braun, U., Garfinkel, S., Holland, D., Muniswamy-Reddy, K., Seltzer, M.: Issues
  in automatic provenance collection.
\newblock Provenance and annotation of data pp. 171--183 (2006)

\bibitem{Chapman:2008:EPS:1376616.1376715}
Chapman, A.P., Jagadish, H.V., Ramanan, P.: Efficient provenance storage.
\newblock In: SIGMOD, pp. 993--1006 (2008)

\bibitem{chebotko2008scientific}
Chebotko, A., Chang, S., Lu, S., Fotouhi, F., Yang, P.: Scientific workflow
  provenance querying with security views.
\newblock In: Web-Age Information Management (WAIM), pp. 349--356 (2008)

\bibitem{cheney12:_princ_of_proven_dagst_semin}
Cheney, J., Finkelstein, A., Lud{\"a}scher, B., Vansummeren, S.: Principles of
  Provenance (Dagstuhl Seminar 12091).
\newblock Dagstuhl Reports \textbf{2}(2), 84--113 (2012).
\newblock \url{http://dx.doi.org/10.4230/DagRep.2.2.84}

\bibitem{cohen2011search}
Cohen-Boulakia, S., Leser, U.: Search, adapt, and reuse: the future of
  scientific workflows.
\newblock ACM SIGMOD Record \textbf{40}(2), 6--16 (2011)

\bibitem{BioMoby}
Consortium, T.B.: Interoperability with {M}oby 1.0---It's better than sharing
  your toothbrush!
\newblock Briefings in Bioinformatics \textbf{9}(3), 220--231 (2008)

\bibitem{curcin2008scientific}
Curcin, V., Ghanem, M.: Scientific workflow systems---can one size fit all?
\newblock In: Biomedical Engineering Conference (CIBEC) (2008)

\bibitem{davidson2010privacy}
Davidson, S., Khanna, S., Roy, S., Boulakia, S.: {Privacy issues in scientific
  workflow provenance}.
\newblock In: Intl.\ Workshop on Workflow Approaches to New Data-centric
  Science (2010)

\bibitem{de2009design}
De~Roure, D., Goble, C., Stevens, R.: The Design and Realisation of the
  myExperiment Virtual Research Environment for Social Sharing of Workflows.
\newblock Future Generation Computer Systems \textbf{25}(5), 561--567 (2009)

\bibitem{dean2008mapreduce}
Dean, J., Ghemawat, S.: MapReduce: Simplified data processing on large
  clusters.
\newblock CACM \textbf{51}(1), 107--113 (2008)

\bibitem{deelman2004pegasus}
Deelman, E., Blythe, J., Gil, Y., Kesselman, C., Mehta, G., Patil, S., Su, M.,
  Vahi, K., Livny, M.: Pegasus: Mapping scientific workflows onto the grid.
\newblock In: Grid Computing, pp. 131--140. Springer (2004)

\bibitem{deelman09:_workf_and_e_scien}
Deelman, E., Gannon, D., Shields, M., Taylor, I.: Workflows and e-Science: An
  overview of workflow system features and capabilities.
\newblock Future Generation Computer Systems \textbf{25}(5), 528--540 (2009)

\bibitem{Deutch:2012:SQL:2109684.2109809}
Deutch, D., Milo, T.: A structural/temporal query language for Business
  Processes.
\newblock J. Comput. Syst. Sci. \textbf{78}(2), 583--609 (2012)

\bibitem{dey2012datalog}
Dey, S., K{\"o}hler, S., Bowers, S., Lud{\"a}scher, B.: Datalog as a Lingua
  Franca for Provenance Querying and Reasoning.
\newblock In: Workshop on the Theory and Practice of Provenance (TaPP) (2012)

\bibitem{dey2011propub}
Dey, S., Zinn, D., Lud\"ascher, B.: {PROPUB: Towards a Declarative Approach for
  Publishing Customized, Policy-Aware Provenance}.
\newblock In: Intl.\ Conf.\ on Scientific and Statistical Database Management
  (SSDBM) (2011)

\bibitem{Dijkman:2009:GMA:1617309.1617317}
Dijkman, R., Dumas, M., Garc\'{\i}a-Ba\~{n}uelos, L.: Graph Matching Algorithms
  for Business Process Model Similarity Search.
\newblock In: Intl Conf.\ on Business Process Management (BPM), pp. 48--63
  (2009)

\bibitem{Dong99maintainingtransitive}
Dong, G., Libkin, L., Su, J., Wong, L.: Maintaining Transitive Closure of
  Graphs in SQL.
\newblock Int. J. Information Technology \textbf{5} (1999)

\bibitem{DBLP:journals/procedia/DouCMMLMH12}
Dou, L., Cao, G., Morris, P.J., Morris, R.A., Lud\"ascher, B., Macklin, J.A.,
  Hanken, J.: Kurator: A Kepler Package for Data Curation Workflows.
\newblock Procedia CS \textbf{9}, 1614--1619 (2012).
\newblock Demo video at \url{http://youtu.be/DEkPbvLsud0}

\bibitem{dou11:_scien_workf_desig}
Dou, L., Zinn, D., McPhillips, T.M., K{\"o}hler, S., Riddle, S., Bowers, S.,
  Lud{\"a}scher, B.: Scientific workflow design 2.0: Demonstrating streaming
  data collections in Kepler.
\newblock In: Intl.\ Conf.\ on Data Engineering (ICDE) (2011)

\bibitem{TamingHeterogeneityPtolemyApproach}
Eker, J., Janneck, J., Lee, E.A., Liu, J., Liu, X., Ludvig, J., Sachs, S.,
  Xiong, Y.: Taming heterogeneity - the Ptolemy approach.
\newblock Proceedings of the IEEE \textbf{91}(1), 127--144 (2003)

\bibitem{ellqvist2009using}
Ellqvist, T., Koop, D., Freire, J., Silva, C., Stromback, L.: Using mediation
  to achieve provenance interoperability.
\newblock In: Services-I, 2009 World Conference on, pp. 291--298. IEEE (2009)

\bibitem{fagin2009clio}
Fagin, R., Haas, L., Hern{\'a}ndez, M., Miller, R., Popa, L., Velegrakis, Y.:
  Clio: Schema mapping creation and data exchange.
\newblock Conceptual Modeling: Foundations and Applications pp. 198--236 (2009)

\bibitem{Fernandez:2000:DSW:765218.765222}
Fern\'{a}ndez, M., Florescu, D., Levy, A., Suciu, D.: Declarative specification
  of Web sites with S.
\newblock The VLDB Journal \textbf{9}(1), 38--55 (2000)

\bibitem{Freire:2006:MRS:2165554.2165557}
Freire, J., Silva, C.T., Callahan, S.P., Santos, E., Scheidegger, C.E., Vo,
  H.T.: Managing rapidly-evolving scientific workflows.
\newblock In: Intl.\ Annotation and Provenance Workshop (IPAW), pp. 10--18
  (2006)

\bibitem{gadelha2011provenance}
Gadelha, L., Mattoso, M., Wilde, M., Foster, I.: Provenance query patterns for
  Many-Task scientific computing.
\newblock Workshop on the Theory and Practice of Provenance. Heraklion, Greece
  pp. 1--6 (2011)

\bibitem{GadelhaJr.:2011:PMS:1967762.1967947}
Gadelha~Jr., L.M.R., Clifford, B., Mattoso, M., Wilde, M., Foster, I.:
  Provenance management in Swift.
\newblock Future Gener. Comput. Syst. \textbf{27}(6), 775--780 (2011)

\bibitem{geilen2003requirements}
Geilen, M., Basten, T.: Requirements on the execution of Kahn process networks.
\newblock Programming Languages and Systems pp. 319--334 (2003)

\bibitem{gil2007wings}
Gil, Y., Ratnakar, V., Deelman, E., Mehta, G., Kim, J.: Wings for Pegasus:
  Creating large-scale scientific applications using semantic representations
  of computational workflows.
\newblock In: National Conference on Artificial Intelligence, vol.~22 (2007)

\bibitem{DBLP:conf/iccS/GoderisBALG07}
Goderis, A., Brooks, C., Altintas, I., Lee, E.A., Goble, C.A.: Composing
  Different Models of Computation in Kepler and Ptolemy II.
\newblock In: Intl.\ Conf.\ on Computational Science (2007)

\bibitem{hellerstein2010declarative}
Hellerstein, J.: The declarative imperative: experiences and conjectures in
  distributed logic.
\newblock SIGMOD Record \textbf{39}(1), 5--19 (2010)

\bibitem{hidders2008dfl}
Hidders, J., Kwasnikowska, N., Sroka, J., Tyszkiewicz, J., Van~den Bussche, J.:
  DFL: A dataflow language based on Petri nets and nested relational calculus.
\newblock Information Systems \textbf{33}(3), 261--284 (2008)

\bibitem{howe2009scientific}
Howe, B., Green-Fishback, H., Maier, D.: Scientific Mashups:
  Runtime-Configurable Data Product Ensembles.
\newblock In: Intl.\ Conf.\ on Scientific and Statistical Database Management
  (SSDBM), pp. 19--36 (2009)

\bibitem{huang2011datalog}
Huang, S., Green, T., Loo, B.: Datalog and emerging applications: an
  interactive tutorial.
\newblock In: SIGMOD, pp. 1213--1216 (2011)

\bibitem{hughes05:_progr_with_arrow}
Hughes, J.: Programming with Arrows.
\newblock In: Intl.\ Summer School on Advanced Functional Programming, LNCS
  3622, pp. 73--129 (2005)

\bibitem{hull2004treating}
Hull, D., Stevens, R., Lord, P., Wroe, C., Goble, C.: Treating ``shimantic
  web'' syndrome with ontologies.
\newblock In: First AKT workshop on Semantic Web Services (2004)

\bibitem{Jin:2011:PER:1929934.1929941}
Jin, R., Ruan, N., Xiang, Y., Wang, H.: Path-tree: An efficient reachability
  indexing scheme for large directed graphs.
\newblock ACM TODS \textbf{36}(1), 7:1--7:44 (2011)

\bibitem{DBLP:conf/ifip/Kahn74}
Kahn, G.: The Semantics of Simple Language for Parallel Programming.
\newblock In: IFIP Congress, pp. 471--475 (1974)

\bibitem{FT}
K{\"o}hler, S., Riddle, S., Zinn, D., McPhillips, T.M., Lud{\"a}scher, B.:
  Improving Workflow Fault Tolerance through Provenance-Based Recovery.
\newblock In: Intl.\ Conf.\ on Scientific and Statistical Database Management
  (SSDBM), pp. 207--224 (2011)

\bibitem{RPQLG}
Koschmieder, A., Leser, U.: Regular path queries on large graphs.
\newblock In: Intl.\ Conf.\ on Scientific and Statistical Database Management
  (SSDBM) (2012)

\bibitem{lee08:_seman_of_dataf_with_firin}
Lee, E.A., Matsikoudis, E.: The Semantics of Dataflow with Firing.
\newblock In: G.~Huet, G.~Plotkin, J.J. L\'evy, Y.~Bertot (eds.) From Semantics
  to Computer Science: Essays in memory of Gilles Kahn (2008)

\bibitem{lee95dataflow}
Lee, E.A., Parks, T.M.: Dataflow Process Networks.
\newblock In: Proceedings of the IEEE, pp. 773--799 (1995)

\bibitem{Li:2011:PBK:1938123.1938150}
Li, G., Feng, J., Zhou, X., Wang, J.: Providing built-in keyword search
  capabilities in RDBMS.
\newblock The VLDB Journal \textbf{20}(1), 1--19 (2011)

\bibitem{lin2009task}
Lin, C., Lu, S., Fei, X., Pai, D., Hua, J.: A task abstraction and mapping
  approach to the shimming problem in scientific workflows.
\newblock In: Services Computing, pp. 284--291. IEEE (2009)

\bibitem{Kepler}
Lud{\"a}scher, B., Altintas, I., Berkley, C., Higgins, D., Jaeger, E., Jones,
  M., Lee, E., Tao, J., Zhao, Y.: Scientific workflow management and the Kepler
  system.
\newblock Concurrency and Computation: Practice \& Experience (CCPE)
  \textbf{18}(10), 1039--1065 (2006)

\bibitem{ludaescher09:_scien_proces_autom_and_workf_manag}
Lud\"ascher, B., Altintas, I., Bowers, S., Cummings, J., Critchlow, T.,
  Deelman, E., Roure, D.D., Freire, J., Goble, C., Jones, M., Klasky, S.,
  McPhillips, T., Podhorszki, N., Silva, C., Taylor, I., Vouk, M.: Scientific
  Process Automation and Workflow Management.
\newblock In: A.~Shoshani, D.~Rotem (eds.) Scientific Data Management. Chapman
  \& Hall/CRC (2009)

\bibitem{ludaescher09:_scien_workf}
Lud\"ascher, B., Bowers, S., McPhillips, T.: Scientific Workflows.
\newblock In: T.~\"Ozsu, L.~Liu (eds.) Encyclopedia of Database Systems.
  Springer (2009)

\bibitem{Ludascher:2009:SWB:1617309.1617315}
Lud\"{a}scher, B., Weske, M., McPhillips, T., Bowers, S.: Scientific Workflows:
  Business as Usual?
\newblock In: Intl.\ Conf.\ on Business Process Management (BPM), pp. 31--47
  (2009)

\bibitem{mcphillips06:_collec_orien_scien_workf_integ}
McPhillips, T., Bowers, S., Lud\"ascher, B.: Collection-Oriented Scientific
  Workflows for Integrating and Analyzing Biological Data.
\newblock In: Intl.\ Workshop on Data Integration in the Life Sciences (DILS)
  (2006)

\bibitem{mcphillips08:_scien_workf_for_mere_mortal}
McPhillips, T., Bowers, S., Zinn, D., Lud\"ascher, B.: Scientific Workflows for
  Mere Mortals.
\newblock Future Generation Computer Systems \textbf{25}(5), 541--551 (2009)

\bibitem{Mendelzon:1995:FRS:219375.219400}
Mendelzon, A.O., Wood, P.T.: Finding Regular Simple Paths in Graph Databases.
\newblock SIAM J. Comput. \textbf{24}(6), 1235--1258 (1995)

\bibitem{5671861}
Missier, P., Ludascher, B., Bowers, S., Dey, S., Sarkar, A., Shrestha, B.,
  Altintas, I., Anand, M., Goble, C.: Linking multiple workflow provenance
  traces for interoperable collaborative science.
\newblock In: Workflows in Support of Large-Scale Science (WORKS), 2010 5th
  Workshop on, pp. 1 --8 (2010)

\bibitem{missier2010linking}
Missier, P., Lud\"ascher, B., Bowers, S., Dey, S., Sarkar, A., Shrestha, B.,
  Altintas, I., Anand, M., Goble, C.: Linking multiple workflow provenance
  traces for interoperable collaborative science.
\newblock In: Workshop on Workflows in Support of Large-Scale Science (WORKS)
  (2010)

\bibitem{Missier:2010:FEL:1739041.1739079}
Missier, P., Paton, N.W., Belhajjame, K.: Fine-grained and efficient lineage
  querying of collection-based workflow provenance.
\newblock In: EDBT, pp. 299--310 (2010)

\bibitem{Missier:2010:TR:1876037.1876076}
Missier, P., Soiland-Reyes, S., Owen, S., Tan, W., Nenadic, A., Dunlop, I.,
  Williams, A., Oinn, T., Goble, C.: Taverna, reloaded.
\newblock In: Intl.\ Conf.\ on Scientific and Statistical Database Management
  (SSDBM), pp. 471--481 (2010)

\bibitem{Moreau:2011:OPM:1967762.1967931}
Moreau, L., Clifford, B., Freire, J., Futrelle, J., Gil, Y., Groth, P.,
  Kwasnikowska, N., Miles, S., Missier, P., Myers, J., Plale, B., Simmhan, Y.,
  Stephan, E., den Bussche, J.V.: The Open Provenance Model core specification
  (v1.1).
\newblock Future Gener. Comput. Syst. \textbf{27}(6), 743--756 (2011)

\bibitem{Moreau:FOPM09}
Moreau, L., Kwasnikowska, N., den Bussche, J.V.: A Formal Account of the Open
  Provenance Model.
\newblock Tech. rep., University of Southampton (2009)

\bibitem{Muniswamy-Reddy:2009:LPS:1855807.1855817}
Muniswamy-Reddy, K.K., Braun, U., Holland, D.A., Macko, P., Maclean, D., Margo,
  D., Seltzer, M., Smogor, R.: Layering in provenance systems.
\newblock In: USENIX (2009)

\bibitem{ngu2008flexible}
Ngu, A., Bowers, S., Haasch, N., McPhillips, T., Critchlow, T.: Flexible
  scientific workflow modeling using frames, templates, and dynamic embedding.
\newblock In: Intl.\ Conf.\ on Scientific and Statistical Database Management
  (SSDBM), pp. 566--572 (2008)

\bibitem{ogasawara:hal-00640431}
Ogasawara, E., De~Oliveira, D., Valduriez, P., Dias, D., Porto, F., Mattoso,
  M.: {An Algebraic Approach for Data-Centric Scientific Workflows}.
\newblock Proceedings of VLDB \textbf{4}(11), 1328--1339 (2011)

\bibitem{Podhorszki:2007:WAP:1273360.1273368}
Podhorszki, N., Lud\"ascher, B., Klasky, S.A.: Workflow automation for
  processing plasma fusion simulation data.
\newblock In: Workflows in Support of Large-Scale Science (WORKS), pp. 35--44
  (2007)

\bibitem{shankar2005integrating}
Shankar, S., Kini, A., DeWitt, D., Naughton, J.: Integrating databases and
  workflow systems.
\newblock ACM SIGMOD Record \textbf{34}(3) (2005)

\bibitem{Tan:2009:BSW:1534121.1534133}
Tan, W., Missier, P., Madduri, R., Foster, I.: Service-Oriented Computing ---
  ICSOC 2008 Workshops.
\newblock chap. Building Scientific Workflow with Taverna and BPEL: A
  Comparative Study in caGrid, pp. 118--129. Springer-Verlag, Berlin,
  Heidelberg (2009)

\bibitem{taylor07:_workf_for_e_scien}
Taylor, I., Deelman, E., Gannon, D., Shields, M. (eds.): Workflows for
  e-Science: Scientific Workflows for Grids.
\newblock Springer (2007)

\bibitem{Tekle:2010:GQT:1836089.1836093}
Tekle, K.T., Gorbovitski, M., Liu, Y.A.: Graph queries through datalog
  optimizations.
\newblock In: Principles and Practice of Declarative Programming (PPDP), pp.
  25--34 (2010)

\bibitem{thain2005distributed}
Thain, D., Tannenbaum, T., Livny, M.: Distributed computing in practice: The
  Condor experience.
\newblock Concurrency and Computation: Practice \& Experience \textbf{17}(2-4),
  323--356 (2005)

\bibitem{thusoo2009hive}
Thusoo, A., Sarma, J., Jain, N., Shao, Z., Chakka, P., Anthony, S., Liu, H.,
  Wyckoff, P., Murthy, R.: Hive: a warehousing solution over a map-reduce
  framework.
\newblock VLDB \textbf{2}(2) (2009)

\bibitem{Turi:2007:TWS:1332478.1333522}
Turi, D., Missier, P., Goble, C., Roure, D.D., Oinn, T.: Taverna Workflows:
  Syntax and Semantics.
\newblock In: Intl.\ Conf.\ on e-Science and Grid Computing (2007)

\bibitem{vrba2009kahn}
Vrba, {\v{Z}}., Halvorsen, P., Griwodz, C., Beskow, P.: Kahn process networks
  are a flexible alternative to MapReduce.
\newblock In: High Performance Computing and Communications (HPCC), pp.
  154--162 (2009)

\bibitem{vrba2010nornir}
Vrba, {\v{Z}}., Halvorsen, P., Griwodz, C., Beskow, P., Espeland, H., Johansen,
  D.: The Nornir run-time system for parallel programs using Kahn process
  networks on multi-core machines—a flexible alternative to MapReduce.
\newblock Journal of Supercomputing pp. 1--27 (2010)

\bibitem{wainer96:_scien_workf_system}
Wainer, J., Weske, M., Vossen, G., Medeiros, C.: Scientific workflow systems.
\newblock In: NSF Workshop on Workflow and Process Automation in Information
  Systems: State-of-the-Art and Future Directions. Athens, GA (1996)

\bibitem{wang12:_early_cloud_exper_with_kepler}
Wang, J., Altintas, I.: Early Cloud Experiences with the Kepler Scientific
  Workflow System.
\newblock Procedia CS \textbf{9}, 1630--1634 (2012)

\bibitem{wang2009kepler+}
Wang, J., Crawl, D., Altintas, I.: Kepler+Hadoop: A general architecture
  facilitating data-intensive applications in scientific workflow systems.
\newblock In: Workshop on Workflows in Support of Large-Scale Science (WORKS)
  (2009)

\bibitem{wieczorek2005scheduling}
Wieczorek, M., Prodan, R., Fahringer, T.: Scheduling of scientific workflows in
  the ASKALON grid environment.
\newblock SIGMOD Record \textbf{34}(3), 56--62 (2005)

\bibitem{wilde2009parallel}
Wilde, M., Foster, I., Iskra, K., Beckman, P., Zhang, Z., Espinosa, A.,
  Hategan, M., Clifford, B., Raicu, I.: Parallel scripting for applications at
  the petascale and beyond.
\newblock IEEE Computer \textbf{42}(11), 50--60 (2009)

\bibitem{wombacher2010data}
Wombacher, A.: Data workflow: a workflow model for continuous data processing
  (2010).
\newblock Centre for Telematics and Information Technology, University of
  Twente

\bibitem{Wood:2012:QLG:2206869.2206879}
Wood, P.T.: Query languages for graph databases.
\newblock SIGMOD Rec. \textbf{41}(1), 50--60 (2012)

\bibitem{Yan:2012:BPM:2109685.2109826}
Yan, Z., Dijkman, R., Grefen, P.: Business process model repositories -
  Framework and survey.
\newblock Inf. Softw. Technol. \textbf{54}(4), 380--395 (2012)

\bibitem{zinn2010xml}
Zinn, D., Bowers, S., Lud\"ascher, B.: XML-based computation for scientific
  workflows.
\newblock In: Intl.\ Conf.\ on Data Engineering (ICDE), pp. 812--815. IEEE
  (2010)

\bibitem{zinn2009scientific}
Zinn, D., Bowers, S., Mc{P}hillips, T., Lud{\"a}scher, B.: Scientific workflow
  design with data assembly lines.
\newblock In: Workshop on Workflows in Support of Large-Scale Science (WORKS)
  (2009)

\bibitem{zinn2009x}
Zinn, D., Bowers, S., Mc{P}hillips, T., Lud\"{a}scher, B.: X-CSR: Dataflow
  Optimization for Distributed XML Process Pipelines.
\newblock In: Intl.\ Conf.\ on Data Engineering (ICDE), pp. 577--580 (2009)

\bibitem{CloudStream}
Zinn, D., Hart, Q., McPhillips, T.M., Lud{\"a}scher, B., Simmhan, Y.,
  Giakkoupis, M., Prasanna, V.K.: Towards Reliable, Performant Workflows for
  Streaming-Applications on Cloud Platforms.
\newblock In: Intl.\ Symposium on Cluster, Cloud and Grid Computing (CCGRID),
  pp. 235--244 (2011)

\bibitem{APG}
Zinn, D., Lud{\"a}scher, B.: Abstract Provenance Graphs: Anticipating and
  Exploiting Schema-Level Data Provenance.
\newblock In: Intl.\ Provenance and Annotation Workshop (IPAW), pp. 206--215
  (2010)

\end{thebibliography}

\end{document}